\documentclass[twoside,12pt]{article}  
\usepackage{CJK}   
\usepackage{indentfirst}  
\usepackage{bm}    
\usepackage{graphicx}  
\usepackage{amsmath}
\usepackage{amsfonts}
\usepackage{amssymb}
\usepackage{mathrsfs}
\usepackage{dcolumn}
\usepackage{epsfig}%

\begin{document}    




\begin{center}
\LARGE\bf Review of cavity optomechanical cooling $^{*}$   
\end{center}

\footnotetext{\hspace*{-.45cm}\footnotesize $^*$We thank Yi-Wen Hu and Xingsheng Luan for helpful discussions. Project supported by 973 program (Grant No. 2013CB328704), National Natural Science Foundation of China (Grant Nos. 11004003, 11222440, and 11121091), and RFDPH (No. 20120001110068). Y.-C. Liu acknowledges the support from the
Scholarship Award for Excellent Doctoral Student granted by Ministry of Education.}
\footnotetext{\hspace*{-.45cm}\footnotesize $^\dag$E-mail:ycliu@pku.edu.cn}
\footnotetext{\hspace*{-.45cm}\footnotesize $^\ddag$URL: www.phy.pku.edu.cn/$\sim$yfxiao/index.html}

\begin{center}
\rm Yong-Chun Liu$^{\rm a)\dagger}$, \ \ Yu-Wen Hu$^{\rm a)}$, \ \ Chee Wei Wong$^{\rm b)}$ \ and  \ Yun-Feng Xiao$^{\rm a)\ddag}$
\end{center}

\begin{center}
\begin{footnotesize} \sl
${}^{\rm a)}$State Key Laboratory for Mesoscopic Physics and School of Physics, Peking University, Beijing 100871, China \\   
${}^{\rm b)}$Optical Nanostructures Laboratory, Columbia University, New York, New York 10027, USA
\end{footnotesize}
\end{center}


\vspace*{2mm}

\begin{center}
\begin{minipage}{15.5cm}
\parindent 20pt\footnotesize
Quantum manipulation of macroscopic mechanical systems is of great interest in both
fundamental physics and applications ranging from high-precision
metrology to quantum information processing. A crucial
goal is to cool the mechanical system to its quantum ground state. In this review,
we focus on the cavity optomechanical cooling, which
exploits the cavity enhanced interaction between
optical field and mechanical motion to reduce the thermal noise. Recent remarkable
theoretical and experimental efforts in
this field have taken a major step forward in preparing the motional quantum
ground state of mesoscopic mechanical systems. This review first describes the
quantum theory of cavity optomechanical cooling, including quantum noise
approach and covariance approach; then the up-to-date experimental
progresses are introduced. Finally, new cooling
approaches are discussed along the directions of cooling in the strong
coupling regime and cooling beyond the resolved sideband limit.
\end{minipage}
\end{center}

\begin{center}
\begin{minipage}{15.5cm}
\begin{minipage}[t]{2.3cm}{\bf Keywords:}\end{minipage}
\begin{minipage}[t]{13.1cm}
cavity optomechanics, optomechanical cooling, cavity
cooling, ground state cooling, mechanical resonator
\end{minipage}\par\vglue8pt
{\bf PACS:}
42.50.Wk, 07.10.Cm, 42.50.Lc
\end{minipage}
\end{center}

\tableofcontents

\section{Introduction}  
Optomechanics is an emerging field exploring the interaction between light and
mechanical motion. Such interaction originates from the mechanical effect of
light, i.e., optical force. Radiation pressure force (or scattering force) and optical gradient force (or
dipole force) are two typical categories of optical forces.
The radiation pressure force originates from the fact that
light carries momentum. The momentum transfer from light to a mechanical
object exerts a pressure force on the object. This was noticed dating back to
the 17th century by Kepler, who noted that the dust tails of comets point away
from the sun. In the 1970s, H\"{a}nsch and Schawlow \cite{LaserCooling75}, Wineland
and Dehmelt \cite{LaserCooling75-2} pointed out the possibility of
cooling atoms by using radiation pressure force of a laser. This was subsequently
realized experimentally \cite{LaserCoolingRMP86}, and it has now become an important technique for
manipulating atoms.
The gradient force stems from
the electromagnetic field gradient. The nonuniform field polarizes the
mechanical object in a way that the positively and negatively charged sides of
the dipole experience different forces, leading to nonzero net optical force
acting on the object. It was first demonstrated by Ashkin that
focused laser beams can be used to trap micro- and nano-scale particles
\cite{AshkinPRL78}. This has stimulated the technique of optical tweezers,
which are widely used to manipulate living cells, DNA and bacteria.
There are other kinds of optical forces, for instance, photothermal force (bolometric force) \cite{CooThermalNat04},
which results from the thermalelastic effect.

\begin{figure}[tb]
\centerline{\includegraphics[width=8cm]{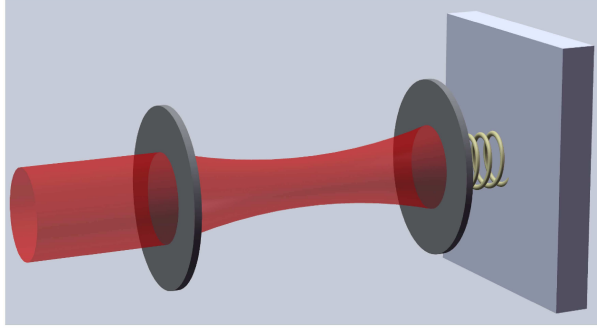}}
\caption{Schematic of a generic optomechanical system, with a laser-driven optical cavity. The left mirror is fixed and the
right mirror is movable.}%
\label{Fig1}%
\end{figure}

The optical forces exerting on macroscopic/mesoscopic mechanical objects are typically very weak.
To overcome this problem, optical cavities are employed, which resonantly enhance the intracavity
light intensity so that the optical forces become pronounced. For example, in a
Fabry-P\'{e}rot (FP) cavity consisting of a fixed mirror and a movable mirror
attached to a spring (Fig. \ref{Fig1}), light is reflected multiple times
between the two mirrors, and thus the cavity field builds up, resulting
largely enhanced optical force exerting on the movable mirror. The study of
this field named as cavity optomechanics was pioneered by Braginsky and
co-workers \cite{Coo1967,Coo1970} with microwave cavities. Later, experiments
in the optical domain demonstrated the optomechanical bistability phenomenon
\cite{WaltherPRL83}, where a macroscopic mirror had two stable equilibrium
positions under the action of cavity-enhanced radiation pressure force.
Further experiments observed optical feedback cooling of mechanical motion
based on precise measurement and active feedback \cite{FBCooPRL98,FBCooPRL99},
and along this line cooling to much lower temperatures was realized later
\cite{FBCooNat06,FBCooPRL07,FBCooPRL07-2}. On the other hand, after the observation of
radiation-pressure induced self-oscillations (parametric instability) in
optical microtoroidal cavities \cite{Amp05PRL,Amp05PRL-2,Amp05OE}, passive
cooling, which uses purely the intrinsic backaction effect of the cavity
optomechanical system, attracted much attention in the past decade
\cite{CooNat06,CooNat06-2,CooPRL06,CooNatPhys08,CooNatPhys09-1,CooNatPhys09-2,CooNatPhys09-3,CooNat10,GSNat11,GSNat11-2,CooPRA11}.
Moreover, recent theoretical and experimental efforts have demonstrated
optomechanically induced transparency
\cite{OMITPRA09,OMITSCI10,OMITNat11,EIAPRA13,OMITPRA13}, optomechanical
storage \cite{StorPRL11}, normal mode splitting
\cite{SCPRL08,SCNat09,SCPRA09,SCNJP10,SCNat11,SCCPB11}, quantum-coherent coupling
between optical modes and mechanical modes \cite{SCNat12,SCNat13} and state
transfer at different optical wavelengths
\cite{STPRA10,STPRL12ydwang,STPRL12tian,STPRL12vitali,ST12NatComm,ST13SCI,ST13PRl}. Various
experimental systems are proposed and investigated, including FP cavities
\cite{CooNat06,CooNat06-2}, whispering-gallery microcavities
\cite{Amp05PRL-2,Sphere09PRL,Sphere07OL,Disk09OE}, microring cavities
\cite{RingNat09}, photonic crystal cavities
\cite{Zipper09Nat,OMcrystal09Nat,LiWongOE10,ZhengWongAPL12}, membranes
\cite{Mem08Nat,Mem11PRAlaw,BuiWongAPL12,hkliPRA12}, nanostrings \cite{Near09NatPhys},
nanorods \cite{LiTangNat08,Nanorod09OE,ZhengWongOE12},\ hybrid plasmonic
structures \cite{ywhuOC13},\ optically levitated particles
\cite{LevitatePNAS10,LevitateNJP10,LevitateNatPhy11,LevitatePRL12,LevitatePRA12,zqyinPRA11,LevitateArXiv13zqyin}%
, cold atoms \cite{CAtomSci08,CAtomNatPhy08,wpzhangBECPRL12} and superconducting
circuits \cite{SuperCNatPhy08}. Recently much attention is also focused
on the studies of single-photon strong optomechanical coupling
\cite{SSC11PRL-1,SSC11PRL-2,SSCCooPRA12,TSC12PRL-1,TSC12PRL-2,yxliuPRA13},
single-photon transport \cite{SPTPRA12,xxrenPRA13,SPTPRA13,SPTarXiv13},
nonlinear quantum optomechanics \cite{NLPRL13-1,NLPRL13-2,ycliuNL13,NLPRL13-4},
quadratic coupling
\cite{Mem08Nat,QuaPRA08,QuaNatPhy10,QuaPRL10,QuaCooPRA10,QuaPRX11,QuaPRL12,QuaPRA12,QuaPRA13}%
, quantum superposition \cite{superPRL11,superPRA11,superPRL12}, entanglement
\cite{EntPRL07,zqyinPRA09,EntPRA11clzou,EntPRA12,EntPRA12-2,EntPRL13tian,EntPRL13wang,EntPRA13yxliu,hkliArXiv13,EntCPB12,EntCPB13,EntCPB13-2}%
, squeezing \cite{SqzPRA09,SqzPRA12,SqzNat12,SqzNat13,SqzPRX13}, decoherence
\cite{DecPRL13}, optomechanical arrays \cite{OMArrPRL13,OMArrNJP12}, quantum hybrid
systems \cite{OptQEDPRL09,OptQEDPRA10,NetworkPRL10,NetworkPRA11,NetworkNJP12},
Brillouin optomechanics \cite{BriNatPhy12,BriNatCom13}, high-precision
measurements \cite{MMSci04,MMNatNano09,MMNatPhoton12,MMMagPRL12,MMSci13,MMPRL13jfdu} and so on.

The rapidly growing interest in cavity optomechanics is a result of the
importance of this subject in both fundamental physics studies and applied
science. On one hand, cavity optomechanics provides a unique platform for the
study of fundamental quantum physics, for example, macroscopic quantum
phenomena, decoherence and quantum-classical boundary. On the other hand,
cavity optomechanics is promising for high-precision measurements of small
forces, masses, displacements and accelerations. Furthermore, cavity
optomechanics provides many useful tools for both classical and quantum
information processing. For instance, optomechanical devices can serve as
storages of information, interfaces between visible light and microwave.
Optomechanical systems also serve as the \textquotedblleft
bridge\textquotedblright\ or \textquotedblleft bus\textquotedblright\ in
hybrid photonic, electronic and spintronic components, providing a routing for
combining different systems to form hybrid quantum devices. A number
of excellent reviews covering various topics has been published in the past
\cite{RevOE07,RevSci08,RevJMO08,RevAAMOP09,RevPhy09,RevNPhot09,RevNPhot10,RevAAMOP10,RevJOSAB10,wpzhangRevFP11,milburnRev11,RevQIP11,RevJPCS11,RevSolid12,RevArxiv12,RevPR12,RevPT12,RevAP13,RevJPB13,kdzhuRev13,RevRMP13,ywhuRev13,zqyinRev13}%
.

As the first crucial step for preparing mechanical quantum states, cooling of
mechanical resonators has been one of the central research interests in the
past decade. Currently, it lacks a comprehensive review on most recent
theoretical and experimental progresses of cavity optomechanical cooling,
especially new cooling approaches for guiding future experiments. In this
review we focus on this issue, addressing the quantum theory, recent
experiments and new directions. The rest of
this paper is organized as follows. In Sec. 2, we present the basic quantum
theory of cavity optomechanical cooling. Starting from the system Hamiltonian,
we introduce the linearization of the interaction. Then the methods for
calculating the cooling rates and cooling limits are shown, including quantum
noise approach and covariance approach. In Sec. 3, we review the
up-to-date experimental progress towards cooling to the quantum ground state.
Recent theoretical approaches for improving the cooling
performance are discussed in Sec. 4. A summary is presented in Sec. 5.

\section{Quantum theory of cavity optomechanical cooling}  
The basic idea of cavity optomechanical cooling is that the optical field
introduces extra damping for the mechanical mode. Qualitatively, such optical
damping is introduced because the optical force induced by the cavity field
reacts with a finite delay time, corresponding to the photon lifetime of the
cavity. Let us take a FP cavity optomechanical system as an
example (Fig. \ref{Fig1}). On one hand, when the movable mirror is at different position, the cavity
field and thus the optical force. exerting on the mirror are also different.
On the other hand, when the position of the movable mirror changes, the
subsequent change of the cavity field requires some time-lag due to the finite
photon lifetime. Therefore, the optical force also depends on the velocity of
the movable mirror. This velocity-dependent optical force
is similar to the friction caused by the intrinsic mechanical damping, and
leads to extra damping (or amplification) of the mechanical motion.
In the classical picture, for an optical damping rate $\Gamma_{\mathrm{opt}}$,
the resulting effective temperature of the mechanical mode being cooled is
$T_{\mathrm{eff}}=\gamma T/(\gamma+\Gamma_{\mathrm{opt}})$, where $\gamma$ is
the intrinsic mechanical damping rate and $T$ is the environmental
temperature. Note that the mechanical mode is selectively cooled, i. e., only
the mode of interest is cooled while the bulk temperature of the mechanical
object keeps unchanged. For the FP cavity case, the mechanical
mode of interest is the center-of-mass motion of the movable mirror.

The above classical description of cavity optomechanical cooling is not accurate in
some cases. For example, it does not predict cooling limits
\cite{PRL07-1,PRL07-2,PRA08}. To accurately model the cooling process, in the
following we provide full quantum theory of cavity optomechanical cooling.
Here both the cavity field and the mechanical oscillation are described as
quantized bosonic fields. Starting from the system Hamiltonian and taking the
dissipations into consideration, we can write down the quantum Langevin
equations and master equation to describe the system dynamics. For different
parameter regimes, cooling performance can be obtained using different approaches.

\subsection{System Hamiltonian and linearization}  
Let us consider a generic cavity optomechanical system with a single optical cavity mode
coupled to a mechanical mode, which is canonically modeled as a
FP cavity with one fixed mirror and one movable mirror mounted on
a spring (Fig. \ref{Fig1}). The system Hamiltonian is given by
\begin{equation}
H=H_{\mathrm{free}}+H_{\mathrm{int}}+H_{\mathrm{drive}}.\label{H}%
\end{equation}

The first term ($H_{\mathrm{free}}$) is the free Hamiltonian of the optical and
mechanical modes, described by
\begin{equation}
H_{\mathrm{free}}={\omega_{\mathrm{c}}a^{\dag}a}+{\omega_{\mathrm{m}}b^{\dag
}b.}%
\end{equation}
Here both of the optical and the mechanical modes are represented by quantum
harmonic oscillators, where ${a}$ (${a^{\dag}}$) is the bosonic annihilation
(creation) operator of the optical cavity mode, ${b}$ (${b^{\dag}}$) is the
bosonic annihilation (creation) operator of the mechanical mode, and ${\omega
}_{\mathrm{c}}$ (${\omega_{\mathrm{m}}}$) is the corresponding angular
resonance frequency. The commutation relations satisfy $[a,{a^{\dag}}]=1$ and
$[b,{b^{\dag}}]=1$. The displacement operator of the mechanical mode is given
by $x=x_{\mathrm{ZPF}}{{({b^{\dag}+b})}}$, where $x_{\mathrm{ZPF}}=\sqrt
{\hbar/(2m_{\mathrm{eff}}{\omega_{\mathrm{m}}})}$ is the zero-point
fluctuation, with $m_{\mathrm{eff}}$ being the effective mass of the
mechanical mode.

The second term of Eq. (\ref{H}) ($H_{\mathrm{int}}$) describes the
optomechanical interaction between the optical mode and the mechanical mode,
which is written as
\begin{equation}
H_{\mathrm{int}}={ga^{\dag}a{({b^{\dag}+b})},}%
\end{equation}
where $g=[\partial{\omega_{\mathrm{c}}(x)/}\partial x]x_{\mathrm{ZPF}}$
represents the single-photon optomechanical coupling strength. This
Hamiltonian can be obtained by simply considering that the cavity resonance
frequency is modulated by the mechanical position and using Taylor expansion
${\omega_{\mathrm{c}}(x)=\omega_{\mathrm{c}}+x}\partial{\omega_{\mathrm{c}%
}(x)/}\partial x+\mathcal{O}(x)\simeq{\omega_{\mathrm{c}}+g{({b^{\dag}+b})}}$.
A more rigorous and detailed derivation of this Hamiltonian can
be found in Law's paper \cite{LawPRA95}. Note that we focus on the radiation pressure
force and the optical gradient force. For the photothermal force, the Hamiltonian
can be found in \cite{ThermalPRA12}.

The last term of Eq. (\ref{H}) ($H_{\mathrm{drive}}$) describes the optical
driving of the system. Assume that the system is excited through a coherent
continuous-wave laser, and then the Hamiltonian is given by
\begin{equation}
H_{\mathrm{drive}}={\Omega}^{\ast}e^{i{\omega}_{\mathrm{in}}t}{a}+{\Omega
}e^{-i{\omega}_{\mathrm{in}}t}{{a^{\dag}}}.
\end{equation}
Here ${\omega}_{\mathrm{in}}$ is the input laser frequency and ${\Omega=}%
\sqrt{\kappa_{\mathrm{ex}}P/(\hbar\omega)}e^{i\phi}$ denotes the driving
strength, where $P$ is the input laser power, $\phi$ is the initial phase of
the input laser and $\kappa_{\mathrm{ex}}$ is the input-cavity coupling rate.

In the frame rotating at the input laser frequency ${\omega}_{\mathrm{in}}$,
the system Hamiltonian is transformed to
\begin{equation}
H={-\Delta a^{\dag}a}+{\omega_{\mathrm{m}}b^{\dag}b}+{ga^{\dag}a{({b^{\dag}%
+b})}}+{{{(}\Omega}^{\ast}a}+{{{\Omega}a^{\dag}),}}%
\end{equation}
where $\Delta=\omega_{\mathrm{in}}-\omega_{\mathrm{c}}$ is the input-cavity
detuning. The quantum Langevin equations are given by
\begin{eqnarray}
& \dot{a}=\left(  i{\Delta}-\frac{\kappa}{2}\right)  a-i{{g}a{(b+b^{\dag
})-i\Omega-}}\sqrt{\kappa_{\mathrm{ex}}}a_{\mathrm{in,ex}}-\sqrt
{\kappa_{\mathrm{0}}}a_{\mathrm{in,0}}{,}\\
& \dot{b}=\left(  -i{\omega_{\mathrm{m}}}-\frac{\gamma}{2}\right)
b-i{{g}a^{\dag}a-}\sqrt{\gamma}b_{\mathrm{in}},
\end{eqnarray} 
where $\kappa_{\mathrm{0}}$ is the intrinsic cavity dissipation rate;
$\kappa=\kappa_{\mathrm{0}}+\kappa_{\mathrm{ex}}$ is the total cavity
dissipation rate; $\gamma$ is the dissipation rate of the mechanical mode;
$a_{\mathrm{in,0}}$, $a_{\mathrm{in,ex}}$ and $b_{\mathrm{in}}$ are the noise
operators associated with the intrinsic cavity dissipation, external cavity
dissipation (input-cavity coupling) and mechanical dissipation. The
correlations for these noise operators are given by
\begin{eqnarray}
& \langle a_{\mathrm{in,0}}(t)a_{\mathrm{in,0}}^{\dag}(t^{\prime})\rangle
=\langle a_{\mathrm{in,ex}}(t)a_{\mathrm{in,ex}}^{\dag}(t^{\prime}%
)\rangle=\delta(t-t^{\prime}),\label{EqCorr1}\\
& \langle a_{\mathrm{in,0}}^{\dag}(t)a_{\mathrm{in,0}}(t^{\prime})\rangle
=\langle a_{\mathrm{in,ex}}^{\dag}(t)a_{\mathrm{in,ex}}(t^{\prime}%
)\rangle=0,\label{EqCorr2}\\
& \langle b_{\mathrm{in}}(t)b_{\mathrm{in}}^{\dag}(t^{\prime})\rangle
=(n_{\mathrm{th}}+1)\delta(t-t^{\prime}),\\
& \langle b_{\mathrm{in}}^{\dag}(t)b_{\mathrm{in}}(t^{\prime})\rangle
=n_{\mathrm{th}}\delta(t-t^{\prime}).
\end{eqnarray}  
Here $n_{\mathrm{th}}$ is the thermal phonon number given by
\begin{equation}
n_{\mathrm{th}}=\left(  e^{\frac{\hbar{\omega_{\mathrm{m}}}}{k_{\mathrm{B}}T}%
}-1\right)  ^{-1},
\end{equation}
where $T$ is the environmental temperature and $k_{\mathrm{B}}$ is Boltzmann constant.
Note that for microwaves the thermal occupations should also be included in Eq. (\ref{EqCorr1})
and (\ref{EqCorr2}). Here we focus on optical frequencies and thus the thermal photon number is negligible.

Coherent laser input results in the displacements of both the optical and
mechanical harmonic oscillators. For convenience, a displacement
transformation is applied, i. e., ${a\rightarrow a}_{1}+\alpha$, $b\rightarrow
b_{1}+\beta$, where $\alpha$ and $\beta$ are $c$-numbers, denoting the
displacements of the optical and mechanical modes; ${a}_{1}$ and $b_{1}$ are
the displaced operators, representing the quantum fluctuations of the optical
and mechanical modes around their classical values. By separating the
classical and quantum components, the quantum Langevin equations are rewritten
as
\begin{eqnarray}
& {\dot{\alpha}}   =\left(  i{\Delta}^{\prime}-\frac{\kappa}{2}\right)
{\alpha}-i{{\Omega},}\label{alpha}\\
& \dot{\beta}   =\left(  -i{\omega_{\mathrm{m}}}-\frac{\gamma}{2}\right)
{\beta}-i{g}\left\vert {\alpha}\right\vert ^{2},\label{beta}\\
& \dot{a}_{1}   =\left(  i{\Delta}^{\prime}-\frac{\kappa}{2}\right)
a_{1}-{{{i{g}\alpha{(b_{1}}}+{{b_{1}^{\dag})}}-i{{g}a}}}_{1}{(b}_{1}%
+{{{{b_{1}^{\dag})}}-}\sqrt{\kappa_{\mathrm{ex}}}a_{\mathrm{in,ex}}%
-\sqrt{\kappa_{\mathrm{0}}}a_{\mathrm{in,0}},}\label{a1}\\
& \dot{b}_{1}   =\left(  -i{\omega_{\mathrm{m}}}-\frac{\gamma}{2}\right)
b_{1}-ig\left(  {\alpha}^{\ast}{a_{1}+{\alpha}a_{1}^{\dag}}\right)
-i{{g}a_{1}^{\dag}a}_{1}-\sqrt{\gamma}b_{\mathrm{in}},\label{a2}%
\end{eqnarray}  
where the optomechanical-coupling modified detuning ${\Delta}^{\prime}%
=\Delta-{{g({\beta}}}+{\beta}^{\ast}{)}$. Under strong driving condition, the
classical components dominate and the nonlinear terms ${{i{{g}a}}}_{1}{(b}%
_{1}+{{{{b_{1}^{\dag})}}}}$ and $i{{g}a_{1}^{\dag}a}_{1}$ in Eqs. (\ref{a1})
and (\ref{a2}) can be neglected, respectively. Then we obtain the linearized
quantum Langevin equations for $a_{1}$ and $b_{1}$, and the corresponding Hamiltonian
is given by
\begin{equation}
H_{L}=-\Delta^{\prime}{a_{1}^{\dag}a}_{1}+{\omega_{\mathrm{m}}{b_{1}^{\dag}%
b}_{1}+(Ga_{1}^{\dag}+G^{\ast}a_{1})(b_{1}+b_{1}^{\dag}}),\label{HL}%
\end{equation}
where $G={\alpha g}$ is the coherent intracavity field enhanced optomechanical
coupling strength.

\begin{figure}[tb]
\centerline{\includegraphics[width=8cm]{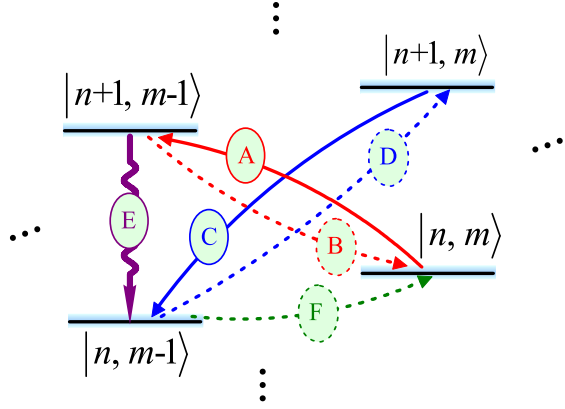}}
\caption{Level diagram of the linearized Hamiltonian (\ref{HL}). $\left\vert
n,m\right\rangle $ denotes the state of $n$ photons and $m$ phonons in the
displaced frame. The solid (dashed) curves with arrows correspond to the
cooling (heating) processes. See text for details.
Figures reproduced with permission from Ref. \cite{ycliuDC13} \copyright \ 2013 American Physical
Society (APS).}%
\label{Fig2}%
\end{figure}

Initially the phonons are in a thermal equilibrium state and the thermal phonon number
is $n_{\mathrm{th}}$. Then the interaction between the photons and the phonons, as
described by the last term in Eq. (\ref{HL}), leads to the modification of the phonon number.
Figure \ref{Fig2} displays the level diagram and the coupling routes among
different states \cite{ycliuDC13}, where $|n,m\rangle$ represents the number state with $n$ ($m$) being the photon (phonon) number in the
displaced frame. Denoted by the dashed curves, there are
three kinds of heating processes: swap heating ($B$), quantum backaction
heating ($D$) and thermal heating ($F$). Thermal heating is an incoherent
process arising from the interaction between the mechanical object and the environment.
Swap heating and quantum backaction heating are the accompanying effect when radiation pressure
is utilized to cool the mechanical motion, corresponding to the coherent
interaction processes ${a_{1}b_{1}^{\dag}}$ and ${a_{1}^{\dag}b_{1}^{\dag}}$,
respectively. Swap heating emerges when the system is in the strong coupling
regime which enables reversible energy exchange between photons and phonons.
Meanwhile, quantum backaction heating can pose a fundamental limit for
backaction cooling. The solid curves ($A$, $C$ and $E$) illustrate cooling
processes associated with energy swapping, counter-rotating-wave interaction
and cavity dissipation. In the following we derive the
cooling rate and cooling limits by taking all the above processes.

\subsection{Quantum noise approach}

In the weak coupling regime, the optomechanical cooling can be analyzed using the
perturbation theory, where optomechanical coupling is regarded as a
perturbation to the optical field. The power spectrum of the optical force
exerting on the mechanical motion $S_{FF}\left(  {\omega}\right)  $ is
calculated with the absence of coupling to the mechanical resonator. Then the
cooling (heating) rate is proportional to $S_{FF}\left(  \pm{{\omega
_{\mathrm{m}}}}\right)  $, corresponding to the ability for absorbing
(emitting) a phonon by the intracavity field.

From Eq. (\ref{HL}) we obtain the optical force acting on the mechanical
motion, described by $F=-({G}^{\ast}{a}_{1}{+{G}a_{1}^{\dag})}/x_{\mathrm{ZPF}}$. The
quantum noise spectrum of the optical force is given by the Fourier transformation
of the autocorrelation function $S_{FF}(\omega)\equiv\int \langle F(t)F(0)\rangle e^{i\omega
t} dt$. The calculation is best performed in
the frequency domain. In the absence of the optomechanical coupling, from Eq.
(\ref{a1}) we obtain
\begin{equation}
-i{\omega{\tilde{a}}}_{1}{(\omega)}=\left(  i\Delta^{\prime}-\frac{\kappa}%
{2}\right)  {{\tilde{a}}}_{1}{(\omega){-}\sqrt{\kappa_{\mathrm{ex}}}\tilde
{a}_{\mathrm{in,ex}}{(\omega)-}\sqrt{\kappa_{\mathrm{0}}}\tilde{a}%
_{\mathrm{in,0}}{(\omega)},}%
\end{equation}
which yields
\begin{equation}
{{\tilde{a}}}_{1}{(\omega)}=\frac{{\sqrt{\kappa}\tilde{a}_{\mathrm{in}%
}{(\omega)}}}{i({\omega}+\Delta^{\prime})-\frac{\kappa}{2}}{,}%
\end{equation}
where ${\tilde{a}_{\mathrm{in}}{(\omega)=}\sqrt{\kappa_{\mathrm{ex}}/\kappa
}\tilde{a}_{\mathrm{in,ex}}{(\omega)+}\sqrt{\kappa_{\mathrm{0}}/\kappa}%
\tilde{a}_{\mathrm{in,0}}{(\omega)}}$. Using $F{\left(  {\omega}\right)
}=-[{G}^{\ast}{{\tilde{a}}}_{1}{(\omega)+{G}\tilde{a}_{1}^{\dag}{(\omega)}%
]}/x_{\mathrm{ZPF}}$, the spectral density of the optical force is obtained as
\begin{equation}
S_{FF}\left(  {\omega}\right)  =\frac{\kappa\left\vert {G}\chi\left(  {\omega
}\right)  \right\vert ^{2}}{x_{\mathrm{ZPF}}^{2}}=\frac{\left\vert
{G}\right\vert ^{2}}{x_{\mathrm{ZPF}}^{2}}\frac{\kappa}{\left[  {\omega
+\Delta^{\prime}}\right]  ^{2}+\frac{\kappa^{2}}{4}}.
\end{equation}
The rate for absorbing and emitting a phonon by the cavity field are
respectively given by
\begin{equation}
A_{\mp}=S_{FF}\left(  \pm{\omega_{\mathrm{m}}}\right)  x_{\mathrm{ZPF}}%
^{2}=\frac{\left\vert {G}\right\vert ^{2}\kappa}{\left[  {\omega_{\mathrm{m}%
}\pm\Delta^{\prime}}\right]  ^{2}+\frac{\kappa^{2}}{4}}.
\end{equation}

We can also derive the spectral density of the mechanical mode $S_{bb}\left(
{\omega}\right)  $ by considering the full equations (Eq. (\ref{a1}) and (\ref{a2}))
\begin{eqnarray}
& -i{\omega{\tilde{a}}}_{1}{(\omega)}   =\left(  i\Delta^{\prime}-\frac{\kappa
}{2}\right)  {{\tilde{a}}}_{1}{(\omega)-iG{{[{{{\tilde{b}_{1}^{\dag}}}%
(\omega)+\tilde{b}}}}}_{1}{{{{(\omega})]}-}\sqrt{\kappa_{\mathrm{ex}}}%
\tilde{a}_{\mathrm{in,ex}}{(\omega)-}\sqrt{\kappa_{\mathrm{0}}}\tilde
{a}_{\mathrm{in,0}}{(\omega)},}\label{a1omega}\\
& -i{\omega}\tilde{b}_{1}({\omega})   =\left(  -i{\omega_{\mathrm{m}}}%
-\frac{\gamma}{2}\right)  \tilde{b}_{1}({\omega})-i{[G}^{\ast}{{\tilde{a}}%
}_{1}{(\omega)+G\tilde{a}_{1}^{\dag}{(\omega)]}}-{\sqrt{\gamma}\tilde
{b}_{\mathrm{in}}(\omega),}  
\end{eqnarray}
from which we obtain
\begin{equation}
\tilde{b}_{1}({\omega
})\simeq\frac{{\sqrt{\gamma}\tilde{b}_{\mathrm{in}}\left(  {\omega}\right)
-i\sqrt{\kappa}}\left\{  {G^{\ast}}\chi\left(  {\omega}\right)  {\tilde
{a}_{\mathrm{in}}{(\omega)}}+{G}\chi^{\ast}\left(  -{\omega}\right)
{\tilde{a}_{\mathrm{in}}^{\dag}{(\omega)}}\right\}  }{i{\omega}-i\left(
{\omega_{\mathrm{m}}+}\Sigma\left(  {\omega}\right)  \right)  -\frac{\gamma
}{2}}.\label{bw}%
\end{equation}
where
\begin{eqnarray}
& \Sigma\left(  {\omega}\right)     =-i\left\vert {G}\right\vert ^{2}\left[
\chi({\omega)}-\chi^{\ast}(-{\omega)}\right]  ,\\
& \chi({\omega})   =\frac{1}{-i({\omega+\Delta^{\prime})}+\frac{\kappa}{2}}.
\end{eqnarray}   
In the second step of the derivation we have neglected the terms containing $\tilde{b}_1^{\dag}(\omega)$, which is negligible near
$\omega=\omega_m$. Here $\Sigma\left(  {\omega}\right)  $ represents the optomechanical self
energy and $\chi({\omega})$ is the response function of the cavity mode. It shows
that the optomechanical coupling leads to the modification of both the mechanical
resonance frequency and the mechanical damping rate, which are termed as optical
spring effect and optical damping effect, respectively. The frequency shift $\delta
{\omega_{\mathrm{m}}}$ and the extra damping $\Gamma_{\mathrm{opt}}$ are given by
\begin{eqnarray}
& \delta{\omega_{\mathrm{m}}}   {=\Re}\Sigma\left(
{\omega_{\mathrm{m}}}\right)  =\left\vert {G}\right\vert ^{2}\Im%
\left[  \frac{1}{-i({\omega_{\mathrm{m}}+\Delta^{\prime})}+\frac{\kappa}{2}%
}{-}\frac{1}{-i({\omega_{\mathrm{m}}-\Delta^{\prime})}+\frac{\kappa}{2}%
}\right]  ,\\
& \Gamma{_{\mathrm{opt}}}   {=-2\Im}\Sigma\left(  {\omega
_{\mathrm{m}}}\right)  =2\left\vert {G}\right\vert ^{2}\Re%
\left[  \frac{1}{-i({\omega_{\mathrm{m}}+\Delta^{\prime})}+\frac{\kappa}{2}%
}{-}\frac{1}{-i({\omega_{\mathrm{m}}-\Delta^{\prime})}+\frac{\kappa}{2}%
}\right]  .
\end{eqnarray}  
The spectral density of the mechanical mode is given by
\begin{equation}
S_{bb}\left(  {\omega}\right)  =%
{\displaystyle\int\nolimits_{-\infty}^{\infty}}
\tilde{b}_{1}^{\dag}\left(  {\omega}\right)  \tilde{b}_{1}\left(  {\omega
}^{\prime}\right)  d{\omega}^{\prime}=\frac{\gamma n_{\mathrm{th}}%
+\kappa\left\vert {G}\chi\left(  -{\omega}\right)  \right\vert ^{2}%
}{\left\vert i{\omega}-i\left(  {\omega_{\mathrm{m}}+\Sigma(\omega)}\right)
-\frac{\gamma}{2}\right\vert ^{2}}.
\end{equation}

\begin{figure}[tb]
\centerline{\includegraphics[width=8cm]{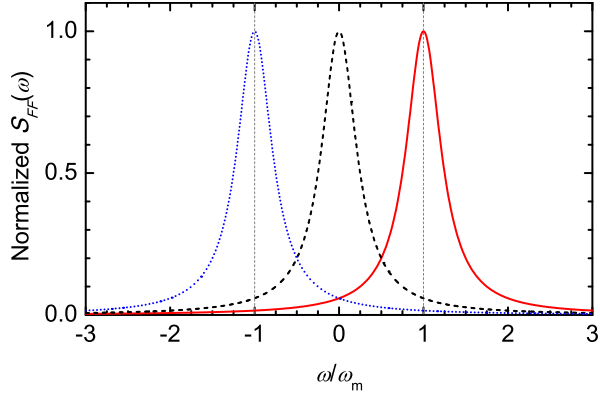}}
\caption{Normalized optical force spectrum $S_{FF}\left(  {\omega}\right)  $ for
$\Delta^{\prime}=-\omega_{\mathrm{m}}$ (red solid curve), $0$ (black
dashed curve), and $\omega_{\mathrm{m}}$ (blue dotted curve). The dashed vertical lines
denotes $\omega/\omega_{\mathrm{m}}=\pm1$. Here $\kappa=0.5\omega_{\mathrm{m}}$.}%
\label{Fig3}%
\end{figure}

Figure \ref{Fig3} plots the normalized $S_{FF}(\omega)$ for different laser detunings.
It shows that red detuning leads to $A_{-}>A_{+}$, corresponding to cooling. In this case,
typically the optical damping rate $\Gamma_{\mathrm{opt}}=A_{-}-A_{+}$ is much
larger than the intrinsic mechanical damping rate, and then the cooling limit
is obtained as
\begin{equation}
n_{\mathrm{f}}=\frac{\gamma n_{\mathrm{th}}+A_{+}}{\Gamma_{\mathrm{opt}}}.
\end{equation}
Note that $n_{\mathrm{f}}^{\mathrm{c}}=\gamma n_{\mathrm{th}}/\Gamma
_{\mathrm{opt}}$ is the classical cooling limit while $n_{\mathrm{f}%
}^{\mathrm{q}}=A_{+}/\Gamma_{\mathrm{opt}}$ corresponds to the fundamental
quantum limit, as the heating rate $A_{+}$ originates from the quantum backaction.
This fundamental limit can be simplified as
\begin{equation}
n_{\mathrm{f}}^{\mathrm{q}}=\frac{4\left(  {\omega_{\mathrm{m}}+\Delta
^{\prime}}\right)  ^{2}+\kappa^{2}}{-16{\omega_{\mathrm{m}}\Delta^{\prime}}}.
\end{equation}
The minimal cooling limit is given by
\begin{equation}
n_{\mathrm{f,\min}}^{\mathrm{q}}=\frac{1}{2}\left(  \sqrt{1+\frac{\kappa^{2}%
}{4{\omega_{\mathrm{m}}^{2}}}}-1\right)  ,
\end{equation}
obtained when $\Delta^{\prime}=-\sqrt{{\omega_{\mathrm{m}}^{2}}+\kappa^{2}/4}$.

In particular, in the unresolved sideband regime (${\omega_{\mathrm{m}}\ll\kappa}$), the
quantum limit is $n_{\mathrm{f,\min}}^{\mathrm{q}}=\kappa/(4{\omega
_{\mathrm{m}}})$ for ${\Delta^{\prime}}=-\kappa/2$. In this case the minimum
phonon number cannot reach 1, which precludes ground state cooling. In the resolved
sideband limit (${\omega_{\mathrm{m}}\gg\kappa}$), the quantum limit is
simplified as $n_{\mathrm{f,\min}}^{\mathrm{q}}=\kappa^{2}/(16{\omega
_{\mathrm{m}}^{2})}$ for ${\Delta^{\prime}}=-{\omega_{\mathrm{m}}}$. In this
limit ground state can be achieved \cite{PRL07-1,PRL07-2}.

\subsection{Covariance approach}

For the linear regime under strong driving, the mean phonon number can be
computed exactly by employing the quantum master equation and solving a linear
system of differential equations involving all the second-order moments. This
approach holds for both weak and strong coupling regimes.

With the linearized Hamiltonian Eq. (\ref{HL}), the quantum master equation reads
\begin{eqnarray}
&  \dot{\rho}   =i[\rho,H_{L}]+\frac{\kappa}{2}\left(  2{a}_{1}\rho{a_{1}^{\dag
}}-{{a_{1}^{\dag}{a}_{1}\rho-\rho a_{1}^{\dag}{a}_{1}}}\right) \nonumber%
\\
& +\frac{\gamma}{2}(n_{\mathrm{th}}+1)\left(  2{b}_{1}\rho{b_{1}^{\dag}%
-{b_{1}^{\dag}b_{1}\rho-\rho b_{1}^{\dag}b_{1}}}\right)  +\frac{\gamma}%
{2}n_{\mathrm{th}}\left(  2{b_{1}^{\dag}}\rho{b}_{1}-{{b_{1}{b_{1}^{\dag}}%
\rho-\rho b_{1}b_{1}^{\dag}}}\right) \label{Master} ,
\end{eqnarray}  
To calculate the mean phonon number, we need to determine the mean values of
all the second-order moments, $\bar{N}_{a}=\langle{a_{1}^{\dag}a}_{1}\rangle$,
$\bar{N}_{b}=\langle{b_{1}^{\dag}b}_{1}\rangle$, $\langle{a{_{1}^{\dag
}{{{b_{1}}}}}}\rangle$, $\langle{a{_{1}{{{b_{1}}}}}}\rangle$, $\langle
{a{_{1}^{2}}}\rangle$ and $\langle{b{_{1}^{2}}}\rangle$
\cite{ycliuDC13,SCNJP08}, which are determined by a linear system of ordinary
differential equations
\begin{eqnarray}
& \frac{d}{dt}\bar{N}_{a}   =-i\left(  G\langle{a{_{1}^{\dag}{{{b_{1}}}}}%
}\rangle-G^{\ast}\langle{a{_{1}^{\dag}{{{b_{1}}}}}}\rangle^{\ast
}+G\left\langle {a_{1}{{{{b_{1}}}}}}\right\rangle ^{\ast}-G^{\ast}\left\langle
{a_{1}{{{{b_{1}}}}}}\right\rangle \right)  -{\kappa}\bar{N}_{a},\\
& \frac{d}{dt}\bar{N}_{b}   =-i\left(  -G\langle{a{_{1}^{\dag}{{{b_{1}}}}}%
}\rangle+G^{\ast}\langle{a{_{1}^{\dag}{{{b_{1}}}}}}\rangle^{\ast
}+G\left\langle {a_{1}{{{{b_{1}}}}}}\right\rangle ^{\ast}-G^{\ast}\left\langle
{a_{1}{{{{b_{1}}}}}}\right\rangle \right)  -{\gamma}\bar{N}_{b}+{\gamma
}n_{\mathrm{th}},\\
& \frac{d}{dt}\langle{a{_{1}^{\dag}{{{b_{1}}}}}}\rangle   =\left[-i\left(
\Delta^{\prime}+{\omega_{\mathrm{m}}}\right)  -\frac{{\kappa+\gamma}}%
{2}\right]\langle{a{_{1}^{\dag}{{{b_{1}}}}}}\rangle-i\left(  G^{\ast}\bar{N}%
_{a}-G^{\ast}\bar{N}_{b}+G\left\langle {a{_{1}^{2}}}\right\rangle ^{\ast
}-G^{\ast}\left\langle {{{{{b_{1}^{2}}}}}}\right\rangle \right)  ,\\
& \frac{d}{dt}\left\langle {a{_{1}{{{b_{1}}}}}}\right\rangle    =\left[i\left(
\Delta^{\prime}-{\omega_{\mathrm{m}}}\right)  -\frac{{\kappa+\gamma}}%
{2}\right]\left\langle {a{_{1}{{{b_{1}}}}}}\right\rangle -i\left(  G\bar{N}%
_{a}+G\bar{N}_{b}+G+G^{\ast}\left\langle {a{_{1}^{2}}}\right\rangle
+G\left\langle {b{_{1}^{2}}}\right\rangle \right)  ,\\
& \frac{d}{dt}\left\langle {a{_{1}^{2}}}\right\rangle    =\left(
2i\Delta^{\prime}-{\kappa}\right)  \left\langle {a{_{1}^{2}}}\right\rangle
-2iG\left(  \left\langle {a}_{1}{{{{{b_{1}}}}}}\right\rangle +\langle
{a{_{1}^{\dag}{{{b_{1}}}}}}\rangle^{\ast}\right)  ,\\
& \frac{d}{dt}\left\langle {b{_{1}^{2}}}\right\rangle    =\left(
-2i{\omega_{\mathrm{m}}}-{\gamma}\right)  \left\langle {b{_{1}^{2}}%
}\right\rangle -2i\left(  G^{\ast}\left\langle {a}_{1}{{{{{b_{1}}}}}%
}\right\rangle +G\langle{a{_{1}^{\dag}{{{b_{1}}}}}}\rangle\right)  .
\end{eqnarray}  
Note that in the above calculation, cut-off of the density matrix is not
necessary and the solutions are exact.

In the stable regime, which requires $\left\vert G\right\vert ^{2}%
<-(4\Delta^{\prime2}+{\kappa}^{2}){\omega_{\mathrm{m}}/(16}\Delta^{\prime})$
for red detuning $\Delta^{\prime}<0$ \cite{StablePRA87}, the system finally
reaches the steady state, and the derivatives in the above equations all become
zero. Then the second-order moments in the steady state satisfy a set of
algebraic equations. Under the condition $\Delta^{\prime}=-{\omega
_{\mathrm{m}}}$ and cooperativity $C\equiv{4}\left\vert {G}\right\vert
^{2}/({\gamma\kappa})\gg1$, the final phonon occupancy reads \cite{ycliuDC13}
\begin{equation}
\bar{N}_{\mathrm{std}}\simeq\frac{{4}\left\vert {G}\right\vert ^{2}+{\kappa
}^{2}}{{4}\left\vert {G}\right\vert ^{2}\left(  {\kappa+\gamma}\right)
}{\gamma}n_{\mathrm{th}}+\frac{{4\omega_{\mathrm{m}}^{2}}\left(  {\kappa}%
^{2}+8\left\vert {G}\right\vert ^{2}\right)  +{{\kappa}^{2}}\left(  {{\kappa
}^{2}}-8\left\vert {G}\right\vert ^{2}\right)  }{{16\omega_{\mathrm{m}}%
^{2}(4\omega_{\mathrm{m}}^{2}+{\kappa}^{2}-16}\left\vert {G}\right\vert ^{2}%
)}.\label{Nstd_s}%
\end{equation}
Here the first term, being proportional to the environmental thermal phonon
number $n_{\mathrm{th}}$, is the classical cooling limit; the second term,
which does not depend on $n_{\mathrm{th}}$, corresponds to the quantum cooling
limit. This quantum limit originates from the quantum backaction, consisting
of both dissipation quantum backaction related to the cavity dissipation and
interaction quantum backaction associated with the optomechanical interaction.
In the resolved sideband case, Eq. (\ref{Nstd_s}) reduces to
\begin{equation}
\bar{N}_{\mathrm{std}}\simeq\frac{{\gamma(4}\left\vert {G}\right\vert
^{2}+{\small \kappa}^{2}{)}}{{4}\left\vert {G}\right\vert ^{2}\left(
{\kappa+\gamma}\right)  }n_{\mathrm{th}}+\frac{{\kappa}^{2}+8\left\vert
{G}\right\vert ^{2}}{{16(\omega_{\mathrm{m}}^{2}-4}\left\vert {G}\right\vert
^{2})}.\label{Nstd}
\end{equation}
In the weak coupling regime, it further reduces to $\bar{N}_{\mathrm{std}%
}^{\mathrm{wk}}\simeq{\gamma}n_{\mathrm{th}}/(\Gamma+\gamma)+{\kappa}%
^{2}/({16\omega_{\mathrm{m}}^{2})}$ with $\Gamma=4|{G}|^{2}/{\kappa}$. In the strong coupling regime, $\bar
{N}_{\mathrm{std}}^{\mathrm{str}}\simeq{\gamma}n_{\mathrm{th}}/({\kappa
+\gamma)}+|{G}|^{2}/[{2}({\omega_{\mathrm{m}}^{2}-4}|{G}|^{2})]$. In this case
the classical limit is restricted by the cavity dissipation rate $\kappa$,
while the interaction quantum backaction limit suffers from high coupling rate
$|{G}|$.

\begin{figure}[tb]
\centerline{\includegraphics[width=8cm]{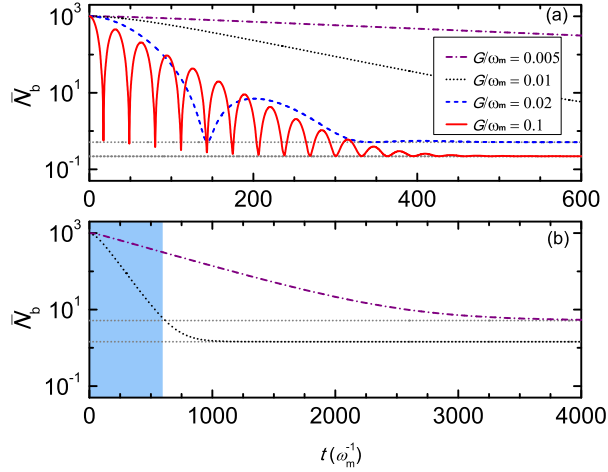}}
\caption{(a) Time evolution of mean phonon number $\bar{N}_{b}$ for
$G/{\omega_{\mathrm{m}}}=0.005$, $0.01$, $0.02$ and $0.1$ (numerical results).
(b) $\bar{N}_{b}$ for $G/{\omega_{\mathrm{m}}}=0.005$ and $0.01$ with a wider
time interval. The shadowed region shows the same time interval with (a).
Other parameters: $n_{\mathrm{th}}=10^{3}$, $\gamma/{\omega}_{\mathrm{m}%
}=10^{-5}$, ${\kappa/\omega_{\mathrm{m}}=0.05}$. The dotted horizontal lines
correspond to the steady-state cooling limits, given by Eq. (\ref{Nstd}).
Figures reproduced with permission from Ref. \cite{ycliuDC13} \copyright \ 2013 APS.
}%
\label{Fig4}%
\end{figure}

To study the cooling dynamics beyond the steady state, the differential
equations need to be solved to obtain the time evolution of the mean
phonon number $\bar{N}_{b}$. For weak coupling, we have $\bar{N}%
_{b}^{\mathrm{wk}}\simeq n_{\mathrm{th}}({\gamma+\Gamma e}^{-\Gamma
t})/({\gamma+\Gamma})+[{\kappa}^{2}/({16\omega_{\mathrm{m}}^{2})](}%
1-{e}^{-\Gamma t}), $ which shows that the mean phonon number decays
exponentially with the cooling rate $\Gamma$. This cooling rate is limited by
the coupling strength, since in the cooling route $A\rightarrow E$ as shown in Fig. \ref{Fig2}, the energy
flow from the mechanical mode to the optical mode (process $A$) is slower than
the cavity dissipation (process $E$).

In the strong coupling regime, the time evolution of the mean phonon
number is described by \cite{ycliuDC13}
\begin{eqnarray}
& \bar{N}_{b}^{\mathrm{str}}   =\bar{N}_{b,1}^{\mathrm{str}}+\bar{N}%
_{b,2}^{\mathrm{str}},\nonumber\\
& \bar{N}_{b,1}^{\mathrm{str}}   \simeq n_{\mathrm{th}}\frac{{\gamma+\frac{{1}%
}{2}e}^{-\frac{{\kappa+\gamma}}{2}t}\left[  {\kappa-\gamma}+({\kappa+\gamma
)}\cos({\omega}_{+}-{\omega}_{-})t\right]  }{{\kappa+\gamma}},\nonumber\\
& \bar{N}_{b,2}^{\mathrm{str}}   \simeq\frac{\left\vert {G}\right\vert
^{2}\left[  1-{e}^{-\frac{{\kappa+\gamma}}{2}t}\cos({\omega}_{+}+{\omega}%
_{-})t\cos({\omega}_{+}-{\omega}_{-})t\right]  }{{2(\omega_{\mathrm{m}}^{2}%
-4}\left\vert {G}\right\vert ^{2})},\label{Nbstr}
\end{eqnarray}
where ${\omega}_{\pm}=\sqrt{{\omega_{\mathrm{m}}^{2}}\pm2|{G}|{\omega
_{\mathrm{m}}}}$ are the normal eigenmode frequencies. The phonon occupancy
exhibits oscillation under an exponentially-decaying envelope and can be
divided into two distinguishable parts $\bar{N}_{b,1}^{\mathrm{str}}$ and
$\bar{N}_{b,2}^{\mathrm{str}}$, where the first part originates from energy
exchange between optical and mechanical modes, and the second part is induced
by quantum backaction. $\bar{N}_{b,1}^{\mathrm{str}}$ reveals Rabi oscillation
with frequency $\sim2|{G}|$, whereas the envelopes have the same exponential
decay rate $\Gamma^{\prime}=({\kappa+\gamma})/2$ regardless of the coupling
strength $|{G}|$. This is because, in the strong coupling regime, the cooling
route $A\rightarrow E$ is subjected to the cavity dissipation (process $E$),
which has slower rate than the energy exchange between phonons and photons
(process $A$). This saturation prevents a higher cooling speed for stronger
coupling. Figures \ref{Fig4}(a) and (b) plot the numerical results based
on the master equation for various $G$. It shows that for weak coupling the
cooling rate increases rapidly as the coupling strength increases, whereas for
strong coupling the envelope decay no longer increases, instead the
oscillation frequency becomes larger.

\section{Recent experimental progresses}

\begin{figure}[tb]
\centerline{\includegraphics[width=8cm]{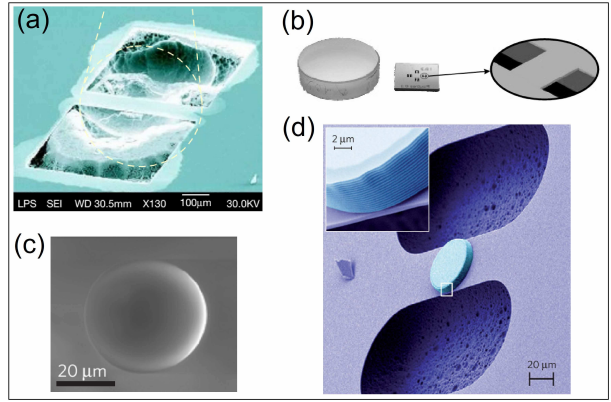}}
\caption{(a) Scanning electron microscope (SEM) image of the cantilever, a doubly clamped free-standing
Bragg mirror (520 $\mu$m long, 120 $\mu$m wide and 2.4 $\mu$m thick) that had been
fabricated by using ultraviolet excimer-laser ablation in combination with a
dry-etching process. (b) Layout of the micromirror optical cavity. The microresonator mirror is etched upon a 1 cm
silicon chip. The coupling mirror of the cavity is a standard low-loss silica mirror.
(c) SEM image of a deformed silica microsphere. (d) SEM image
of the mechanical system formed by a doubly clamped SiN beam. A circular, high-reflectivity
Bragg mirror is used as the end mirror of a FP cavity.
Figures reproduced with permission from: (a) Ref. \cite{CooNat06} \copyright \ 2006 Nature Publishing Group (NPG);
(b) Ref. \cite{CooNat06-2} \copyright \ 2006 NPG; (c) Ref. \cite{CooNatPhys09-2} \copyright \ 2009 NPG;
(d) Ref. \cite{CooNatPhys09-1} \copyright \ 2009 NPG.
}%
\label{Fig5}%
\end{figure}

\begin{figure}[tb]
\centerline{\includegraphics[width=8cm]{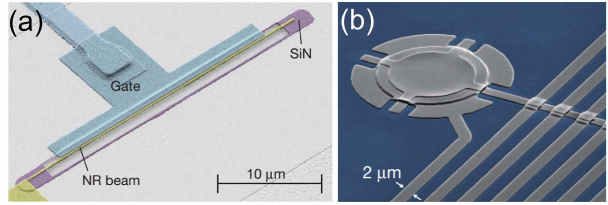}}
\caption{(a) SEM image of the Nb-Al-SiN sample. The nanomechanical resonator is 30 $\mu$m
long, 170 nm wide and 140 nm thick, and is formed of 60 nm of stoichiometric,
high-stress, low-pressure chemical-vapour-deposition SiN and 80 nm of Al.
(b) SEM image of the aluminium (grey) electromechanical circuit
fabricated on a sapphire (blue) substrate. A 15-$\mu$m-diameter membrane is
suspended 50 nm above a lower electrode.
Figures reproduced with permission from: (a) Ref. \cite{CooNat10} \copyright \ 2010 NPG;
(b) Ref. \cite{GSNat11} \copyright \ 2011 NPG.
}%
\label{Fig6}%
\end{figure}

\begin{figure}[tb]
\centerline{\includegraphics[width=8cm]{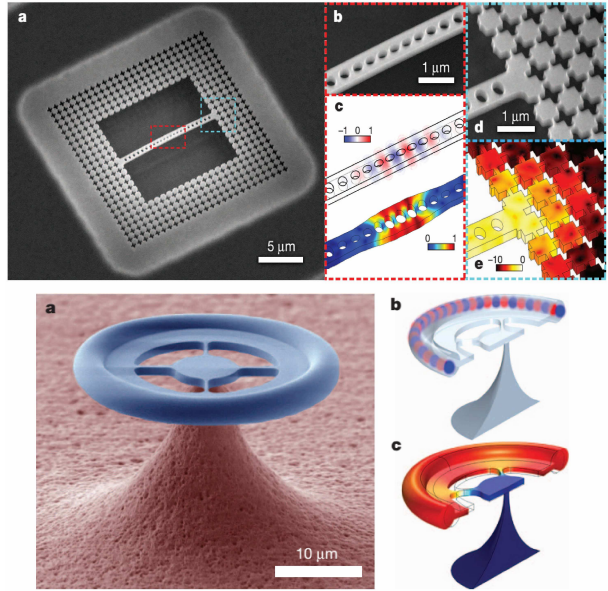}}
\caption{Top panel: Photonic nanocrystal nanobeam cavity with phononic shield. (a)
Scanning electron microscope (SEM) image of the patterned silicon nanobeam and the
external phononic bandgap shield. (b) Enlarged image of the central region of the nanobeam. (c)
Simulation of the optical and mechanical modes. (d) Enlarged image of the nanobeam-shield
interface. (e) Simulation of the localized acoustic resonance at the nanobeam¨Cshield
interface. Bottom panel: Spoke-supported microtoroidal cavity.
(a) SEM image of the spoke-anchored toroidal resonator. (b) Sketch of an
optical whispering gallery mode. (c) Simulation of the fundamental
radial breathing mechanical mode.
Figures reproduced with permission from: Top panel, Ref. \cite{GSNat11-2} \copyright \ 2011 NPG;
Bottom panel, Ref. \cite{SCNat12} \copyright \ 2012 NPG.
}%
\label{Fig7}%
\end{figure}

Pioneering work of cavity optomechanical cooling dates back to 1960th by
Braginsky and coworkers \cite{Coo1967,Coo1970}, where they demonstrated the
modification of mechanical damping rate as a result of the retarded nature of
the cavity-enhanced optical force due to the finite cavity photon lifetime. In
2006, radiation pressure cooling was realized in three groups using different
optomechanical systems, including suspended micromirrors
\cite{CooNat06,CooNat06-2} (Fig. \ref{Fig5} (a) and (b)) and microtoroids \cite{CooPRL06}. Later in
2008, cooling in the resolved sideband regime was achieved \cite{CooNatPhys08}%
. Soon afterwards, with environmental pre-cooling under cryogenic
condition, cooling to only a few phonons was demonstrated
\cite{CooNatPhys09-1,CooNatPhys09-2,CooNatPhys09-3} (Fig. \ref{Fig5} (c) and (d)). Recently, several groups
have cooled the mechanical motion close to the quantum ground state both in
the microwave domain \cite{CooNat10,GSNat11} (Fig. \ref{Fig6}) and in the optical domain
\cite{GSNat11-2,CooPRA11,SCNat12} (Fig. \ref{Fig7}).

More specifically, in Ref. \cite{GSNat11-2}, mean phonon occupancy down to
$0.85$ quanta was achieved, with the ground state occupancy probability
greater than $50\%$ ($P_{g}=0.54$). In this experiment, the cavity
optomechanical system consisted of a photonic crystal nanobeam resonator (Fig. \ref{Fig7} top panel). The
carefully designed periodic patterning of the nanobeam resulted in Bragg
scattering of both optical and acoustic guided waves. At the center of the
nanobeam, a perturbation in the periodicity was introduced, leading to
co-localized optical and mechanical resonances, which are coupled by
optical gradient force. An external acoustic radiation shield consisting of a
two dimensional ``cross'' pattern was designed to minimize the mechanical anchor
damping through phononic bandgap. The optomechanical device was placed in a
continuous-flow He-4 cryostat with pre-cooled environment temperature at $20$ K,
corresponding to about $100$ initial phonon occupancy for the mechanical mode with
resonance frequency as high as $3.68$ GHz. The mechanical Q-factor was
$10^{5}$, corresponding to an intrinsic mechanical damping rate of $35$ kHz.
The optical Q-factor was $4\times10^{5}$, and thus the optical damping rate
was $500$ MHz. By fitting the measured data of mechanical damping effect, the
single-photon optomechanical coupling strength is determined to be $910$ kHz.
With $2000$ intracavity photons, the minimum mean phonon occupancy was
observed to be $0.85\pm0.08$. The cooling was limited for higher drive powers,
which resulted from the increase of the bath temperature due to optical
absorption and the increase of the intrinsic mechanical damping rate induced
by the generation of free carriers through optical absorption.

In Ref. \cite{SCNat12}, a micro-optomechanical system in the form of a
spoke-supported toroidal optical microcavity (Fig. \ref{Fig7} bottom panel) was cooled to $n_{f}=1.7$ quanta.
For such microtoroidal cavity, the supported optical whispering gallery mode
exhibits ultrahigh quality factor exceeding $10^{8}$, and thus the cavity
decay rate reaches $\kappa/2\pi<10$ MHz. The mechanical resonance frequency
for such spoke-anchored toroidal resonator with $31$ $\mathrm{\mu}$\textrm{m}
diameter was $78$ MHz. In a He-3 buffer gas cryostat with $650$ mK
temperature, the mechanical resonator was pre-cooled to $\sim200$ quanta.
Through optomechanical cooling, the final mean phonon occupancy was reduced to
$1.7\pm0.1$. Further cooling was limited by the laser reheating of the sample
and the onset of normal modes. For the latter, cooling in the strong coupling
regime was limited by the swap heating (see the next section). In this
experiment, besides cooling, they also demonstrated the quantum-coherent
coupling between the mechanical mode and the optical mode.

\section{New cooling approaches}

The current cooling approach as shown in Sec. 2 has achieved great successes,
while there are still some major challenges. First, saturation
effect appears in the strong optomechanical coupling regime as a result of
swap heating. Secondly, to achieve ground state cooling, it
requires the resolve sideband condition, which is stringent for many cavity
optomechanical systems. In this section we review recent cooling approaches to
improve the cooling performance along these directions

\subsection{Cooling in the strong coupling regime}

\begin{figure}[tb]
\centerline{\includegraphics[width=8cm]{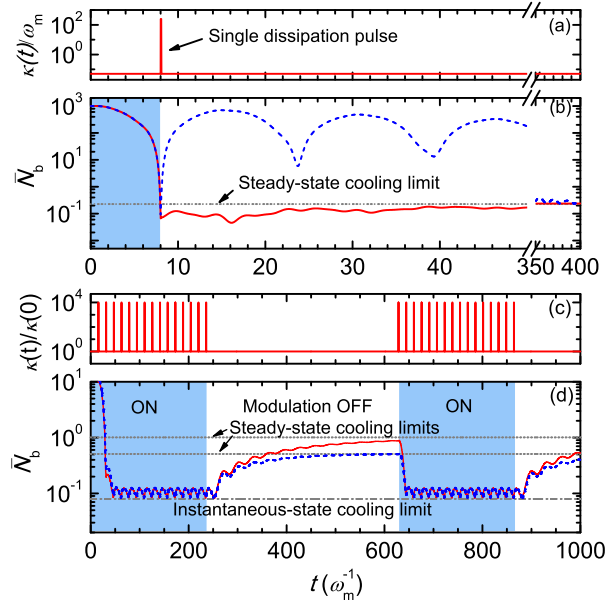}}
\caption{(a) Modulation scheme of the cavity dissipation rate ${\kappa(t)}$
and (b) the corresponding time evolution of mean
phonon number $\bar{N}_{b}$ with (red solid curve) and without (blue dashed
curve) modulation for $G/{\omega_{\mathrm{m}}}=0.2$ and ${\kappa/\omega_{\mathrm{m}}=0.05}$.
(c) Modulation scheme of ${\kappa(t)/\kappa(0)}$ and (d) the
corresponding $\bar{N}_{b}$ for $G/{\omega_{\mathrm{m}}%
}=0.1$, ${\kappa(0)/\omega_{\mathrm{m}}=0.01}$ (red solid curve) and $0.02$
(blue dashed curve). In (d), the two dotted horizontal lines (from top to
bottom) denoting the respective steady-state cooling limits depending on the
cavity decay ${\kappa(0)}$; the dash-dotted line
denotes the instantaneous-state cooling limit independent of ${\kappa(0)}$,
given by Eq. (\ref{Nins}); the \textquotedblleft ON\textquotedblright\ and
\textquotedblleft OFF\textquotedblright\ regions corresponds that the
modulation is turned on and off, respectively; the vertical coordinate range
from $10$ to $10^{3}$ is not shown. Other parameters: $n_{\mathrm{th}}=10^{3}%
$, $\gamma/{\omega}_{\mathrm{m}}=10^{-5}$.
Figures reproduced with permission from Ref. \cite{ycliuDC13} \copyright \ 2013 APS.
}%
\label{Fig8}%
\end{figure}

Recent experiments have reached the regime of strong optomechanical coupling,
which is crucial for coherent quantum optomechanical manipulations. However, as mentioned in Sec. 2,
strongly-coupled optomechanical cooling has predicted only limited
improvement over weak coupling due to the saturation effect of the
steady-state cooling rate. In Ref. \cite{ycliuDC13},
Liu et al. proposed to dynamically tailor the cooling and heating processes by
exploiting the modulation of cavity dissipation. In this proposal, the
internal cavity dissipation is abruptly increased each time when the Rabi
oscillation reaches a minimum-phonon state. At this time the system has transited
from state $|n,m\rangle$ to state $|n+1,m-1\rangle$ (Fig. \ref{Fig2}). Once a strong dissipation
pulse is applied to the cavity so that the process $E$ dominates, the system
will irreversibly transit from state $|n+1,m-1\rangle$ to state $|n,m-1\rangle
$. The dissipation pulse has essentially behaves as a switch to halt the
reversible Rabi oscillation, resulting in the suppression of the swap heating.
Such dissipative cooling is verified in Figs. \ref{Fig8} (a) and (b), which
plot the modulation scheme and the corresponding time evolution of mean phonon
number $\bar{N}_{b}$. At the end of the first half Rabi oscillation
cycle, $t\sim\pi/(2|{G}|)$, a dissipation pulse is applied. After that, the phonon
number reaches and remains near the steady-state limit. Without modulation (blue dashed
curve), the steady-state cooling limit is reached only after $t\simeq
400/\omega_{\mathrm{m}}$; while with the modulation (red solid curve), it only
takes $t\simeq8/\omega_{\mathrm{m}}$ to cool below the same limit, corresponding to 50 times
faster cooling speed.

By periodically modulating the cavity dissipation so as to continuously
suppress the swap heating, the phonon occupancy can be kept below the
steady-state cooling limit. Each time after the dissipation pulse is applied,
the photon number quickly drops to the vacuum state, which equivalently
re-initializes the system. By periodic pulse application, the system will
periodically re-initializes, which keeps the phonon occupancy in an
instantaneous-state cooling limit as verified in Fig. \ref{Fig8} (c) and (d). This limit is given by \cite{ycliuDC13}
\begin{equation}
\bar{N}_{\mathrm{ins}}\simeq\frac{\pi{\gamma}n_{\mathrm{th}}}{4\left\vert
{G}\right\vert }+\frac{\pi^{2}\left\vert {G}\right\vert ^{4}}{{(\omega
_{\mathrm{m}}^{2}-}\left\vert {G}\right\vert ^{2}){(\omega_{\mathrm{m}}^{2}%
-4}\left\vert {G}\right\vert ^{2})}.
\label{Nins}
\end{equation}
Here the first term comes from $\bar{N}_{b,1}^{\mathrm{str}}$ for $t\simeq
\pi/(2|{G}|)$, which shows a $\pi{\kappa/(}4|{G}|)$ times reduction of
classical steady-state cooling limit. The second term of $\sim\pi
^{2}\left\vert {G}\right\vert ^{4}/{\omega_{\mathrm{m}}^{4}}$, obtained from
$\bar{N}_{b,2}^{\mathrm{str}}$ when $t\simeq\pi/{\omega_{\mathrm{m}}} $,
reveals that the second order term of $|{G}|/{\omega_{\mathrm{m}}}$ in quantum
backaction has been removed, leaving only the higher-order terms. It is also demonstrated in Fig. \ref{Fig8} (c) and (d)
that the modulation is switchable. If the modulation is turned
on (\textquotedblleft ON\textquotedblright\ region), the system
will reach the instantaneous-state cooling limit; if the modulation
is turned off (\textquotedblleft OFF\textquotedblright\ region), the system
transits back to the steady-state cooling limit.

Under frequency matching condition $({\omega}_{+}+{\omega}%
_{-})/({\omega}_{+}-{\omega}_{-})=k$ ($k=3,5$...) and $t\simeq
\pi/(2|{G}|)$, the optimized
instantaneous-state cooling limit is obtained as \cite{ycliuDC13}
\begin{equation}
\bar{N}_{\mathrm{ins}}^{\mathrm{opt}}\simeq{\frac{\pi{\kappa}}{4\left\vert
{G}\right\vert }}\left[  \frac{{\gamma}n_{\mathrm{th}}}{{\kappa}}%
+\frac{\left\vert {G}\right\vert ^{2}}{{2(\omega_{\mathrm{m}}^{2}-4}\left\vert
{G}\right\vert ^{2})}\right]  ,
\label{Ninsmat}
\end{equation}
which reduces both the classical and quantum steady-state cooling limits by a
factor of $\pi{\kappa/(}4|{G}|)$. This reduction is significant
when the system is in the deep strong coupling regime. Typically, the cooling limits can be reduced by a few
orders of magnitude. For example, when $G/{\omega_{\mathrm{m}}}=0.3$ and
${\kappa/{\omega_{\mathrm{m}}}=0.003}$, it yields $\bar{N}_{\mathrm{std}}%
=3.4$, while $\bar{N}_{\mathrm{ins}}^{\mathrm{opt}}=0.03$, corresponding to
more than $100$ times of phonon number suppression.

\subsection{Cooling beyond the resolved sideband limit}

To loosen the stringent resolved sideband condition, a few approaches have
been proposed, which can be divided into the following three directions: novel
coupling mechanisms, parameter modulations and hybrid systems.

\subsubsection{Novel coupling mechanisms}

\begin{figure}[tb]
\centerline{\includegraphics[width=8cm]{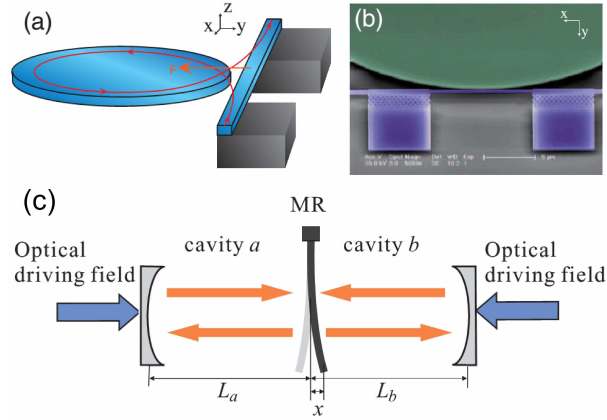}}
\caption{(a) Schematics of the microdisk-waveguide optomechanical system for dissipative coupling.
(b) SEM image of the fabricated device. (c) Schematics of the three-mirror system.
The movable mirror with two perfectly reflecting surfaces
is placed inside a driven cavity with two transmissive fixed mirrors.
Figures reproduced with permission from: (a)-(b) Ref. \cite{LiPRL2009} \copyright \ 2009 APS;
(c) Ref. \cite{PulPRA11-1} \copyright \ 2011 APS.
}%
\label{Fig9}%
\end{figure}

The first direction is searching for novel optomechanical coupling mechanism,
for instance, dissipative coupling, where the mechanical motion couples to the
cavity decay rate instead of the cavity resonance frequency \cite{DCPRL09}. In
the system described in Fig. 1, the displacement of the
mechanical object couples to the cavity resonance frequency $\omega(x)$, which
is sometimes termed dispersive coupling. For the dissipative coupling, the
displacement of the mechanical object couples to the cavity decay rate
$\kappa(x)$. It was predicted \cite{DCPRL09} that this can yield novel cooling
behavior, capable of reaching the quantum ground state without the resolved
sideband limit. The principle is that destructive interference effect occurs
between two noise sources. In the usual dispersive coupling case, the only
source of backaction force noise is the number fluctuations of the cavity
field, leading to Lorentzian noise spectrum. In this dissipative coupling
case, the mechanical oscillator mediates the coupling between the cavity and
the cavity's dissipative bath, and there are two noise sources: one is the
fluctuations of the cavity field and the other is the shot noise associated
with the driving laser. Note that the latter noise process is white, and thus
the interference between these two noises yields a Fano line shape for the
noise spectrum. Such destructive interference allows the cavity to act as an
effective zero-temperature bath at a special detuning, irrespective of the resolved sideband
condition. Recent analysis shows that the quantum cooling limit of cavity
optomechanical system with both dispersive and dissipative couplings can be
optimized \cite{DCPRA13}.

Such dissipative coupling have been observed experimentally in a microdisk
cavity coupled to a waveguide \cite{LiPRL2009} (Fig. \ref{Fig9} (a) and (b)) and many theoretical analyses have been performed in various systems \cite{DCPRL11,DCNJP13,myyanPRA13,DCPRA13gxli}.

\subsubsection{Parameter modulations}

The second direction is to introduce modulations of the system parameters, such as input laser intensity \cite{PulPNAS11,PulPRL11,PulPRL12}, mechanical
resonance frequency \cite{PulPRA11-1} and other parameters \cite{PulPRA12}. In Ref. \cite{PulPRL11,PulPRL12},
optimal control method were introduced, allowing ultra-efficient cooling via
pulsed laser inputs. The idea is to use interference between optical pulses
incident on the system, where a sequence of fast pulses adds a term to the
effective optomechanical interaction Hamiltonian. In the usual continuous
driving case, the interaction term has the form $x_{m}x_{c}$, while pulsed
laser input generates an effective interaction term with the form $p_{m}p_{c}%
$, where $x_{m}$ and $p_{m}$ ($x_{c}$ and $p_{c}$) are the quadrature
operators of the mechanical (optical) mode. By optimizing the pulse duration
time, the total effective interaction is described by the beam-splitter
Hamiltonian $x_{m}x_{c}+p_{m}p_{c}\propto ab^{\dagger}+a^{\dagger}b$. As a
result, the counter-rotating-wave term is eliminated, avoiding the quantum
backaction heating.

In Ref. \cite{PulPRA11-1}, Li et al. propose a ground state cooling scheme by
taking advantage of a mechanical resonator with time-dependent frequency,
using a three-mirror system (Fig. \ref{Fig9} (c)). In
this scheme, strong laser input is used to generate optical spring effect,
where the effective resonance frequency of the mechanical mode is determined
by the optical driving. Fast ground state cooling can be achieved by designing
the trajectory of the effective frequency from the initial time to the final time.

Liao and Law have investigated the cooling performance using chirped-pulse
coupling by modulating the input laser intensity and phase \cite{PulPRA11-2}.
In this scheme, owing to the frequency modulation in chirped pulses, cooling
can be realized without the need for high-precision control of the laser
detuning and pulse areas.

\subsubsection{Hybrid systems}

\begin{figure}[tb]
\centerline{\includegraphics[width=8cm]{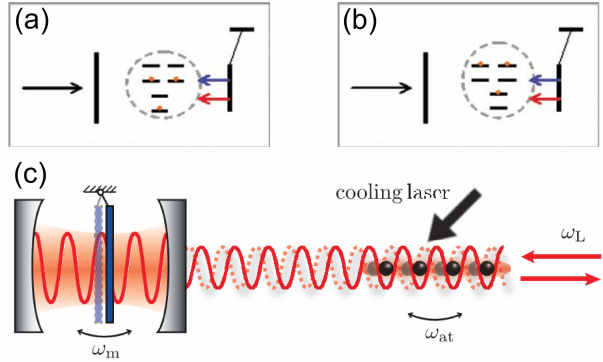}}
\caption{(a) Couple two-level atoms to the optical cavity with a movable mirror. The atoms
are in the ground states. (b) Same as (a) but the atoms are in the excited states. (c)
Micromechanical membrane in a cavity coupled to a distant atomic ensemble.
Figures reproduced with permission from: (a)-(b) Ref. \cite{Atom09PRA} \copyright \ 2009 APS;
(c) Ref. \cite{AtomPRA13} \copyright \ 2013 APS.
}%
\label{Fig10}%
\end{figure}

The third direction is to construct hybrid systems, for example, couple atoms
to the optical cavity. In Ref. \cite{Atom09PRA}, Genes et al. have described a
hybrid system by coupling a two-level ensemble to the cavity optomechanical
system (Fig. \ref{Fig10} (a) and (b)). The two-level ensemble couples to the cavity mode, creating
intracavity narrow bandwidth loss or gain. It induces tailored asymmetric
structure of the cavity noise spectrum interacting with the mechanical mode.
As a result, This allows cooling via inhibition of the Stokes-scattering
process or enhancement of anti-Stokes scattering even for low-finesse optical
cavities. Ground state cooling can be realized without the requirement of
resolved sideband condition, as long as the loss or gain bandwidth of the
two-level ensemble is narrow enough.

Similar hybrid system containing atoms are investigated in Ref.
\cite{AtomPRA13} (Fig. \ref{Fig10} (c)). Here the atomic ensemble is trapped in an optical lattice,
and the center-of-mass motion instead of internal states are considered. The
cavity mode mediates the interaction between the mechanical motion of the
membrane and the center-of-mass motion of the atomic ensemble, leading to the
effective interaction between these two mechanical modes. By laser cooling of
the atomic ensemble, the membrane motion can be cooled via the effective
interaction. This does not require resolved sideband conditions for the
cavity, since the cavity mode acts as an intermediary.

Before conclusion, we would like to make some remarks on the influence of
laser phase noise on cavity optomechanical cooling
\cite{PhasePRA08,PhasePRA09zqyin,PhasePRA09,PhasePRA11,PhasePRA11-2,PhasePRA11-3,PhaseArXiv11,PhaseCPL13}%
, which is a technical factor that limits the cooling process. The phase noise
exists in many lasers, especially in diode lasers \cite{PhaseArXiv11}. The
phase fluctuation of the cooling laser will induce the photon number
fluctuation in the cavity mode. This fluctuation is equivalent to a thermal
bath coupled to the mechanical resonator, and thus it limits the cooling performance.
In Ref. \cite{PhasePRA09zqyin}, Yin has proposed a
double-mode cooling configuration to reduce the influence of the laser phase
noise. A whispering-gallery mode cavity with double optical modes is used, and
the mechanical mode is coupled to both cavity modes, with the resonance
frequency equal to the frequency splitting of the two cavity modes. It is
shown that the phase noise effect can be strongly suppressed when the system
is in the resolved sideband regime.

\section{Summary and outlook}

In summary, we have reviewed the quantum theory, recent experiments and new
directions of cavity optomechanical cooling. Particularly, we summarize the
new cooling scheme along two directions: cooling in the strong coupling regime
and cooling beyond the resolved sideband limit. Novel cooling schemes for
efficiently suppressing the thermal noise are still being explored. These
schemes add complexity to the current experimental systems, and thus efforts
should be taken to demonstrate these schemes in real experiments, which would
be possible in the near future. There are more appealing challenges such as cooling massive mechanical objects up to kilograms and room-temperature ground state cooling of mechanical resonators.

With currently rapid experimental and theoretical advances in cavity
optomechanics, it opens up new avenues to the foundations of quantum physics
and applications. Quantum manipulation of macroscopic/mesoscopic mechanical objects provides
a direct method to test fundamental quantum theory in a hitherto unachieved
parameter regime. For example, micro mechanical structures consist of
typically $10^{14}$ atoms and weigh $10^{-11}$ kilograms, while gravitational
wave detectors comprise more than $10^{20}$ atoms and weigh up to several
kilogram. For applications, cavity optomechanics provides new aspects for
measurement with high precision and for sensing with high sensitivity.
Particularly, cavity optomechanics offers a new architecture for solid-state
realization of quantum information processing. The development of cavity
optomechanical cooling will enable quantum manipulation of mechanical objects
and generation of non-classical mechanical states, which will provide unique
resources for quantum communication and quantum computation.


\end{document}